\documentclass[prl,aps,amssymb,superscriptaddress,twocolumn]{revtex4}
\usepackage{graphicx}
\usepackage{amsmath}
\usepackage{mathtools}
\usepackage{ textcomp }

\begin{document}

\title{Observation of Quantum Oscillations in the Photo-assisted Shot Noise of a Tunnel Junction}

\author{Gabriel Gasse}
\affiliation{D\'{e}partement de Physique, Universit\'{e} de Sherbrooke, Sherbrooke, Qu\'{e}bec, Canada, J1K 2R1}

\author{Lafe Spietz}
\affiliation{National Institute of Standards and Technology, Boulder, Colorado 80305, USA}

\author{Christian Lupien}
\affiliation{D\'{e}partement de Physique, Universit\'{e} de Sherbrooke, Sherbrooke, Qu\'{e}bec, Canada, J1K 2R1}

\author{Bertrand Reulet}
\affiliation{D\'{e}partement de Physique, Universit\'{e} de Sherbrooke, Sherbrooke, Qu\'{e}bec, Canada, J1K 2R1}

\date{\today}
\begin{abstract}
We report measurements of the low frequency current fluctuations of a tunnel junction placed at very low temperature biased by a time-dependent voltage $V(t)=V(1+\cos 2\pi\nu t)$. We observe that the excess noise generated by the ac excitation exhibits quantum oscillations as a function of the dc bias, with a period given by $h\nu/e$ with $e$ the charge of a single electron. This is a direct consequence of the quantum nature of electricity in a normal conductor.\end{abstract}
\maketitle

Quantum oscillations are the most prominent manifestation of the quantum nature of a phenomenon. For instance, the macroscopic phase coherence in superconductors leads to current oscillations in a voltage biased Josephson junction\cite{Josephson}. These engender, in dc transport under irradiation at frequency $\nu$, periodic anomalies vs bias voltage with a period $h\nu/(2e)$, $2e$ being the charge of the Cooper pairs\cite{Shapiro}.  Despite the absence of macroscopic phase coherence, electrons in a normal metal are quantum objects described by a wavefunction with a phase. That phase is random from one electron to the next, yet the effect of the bias voltage is the same for all electrons: it shifts the phase of their wavefunction by the same amount $\phi$. Thus, quantum oscillations with a period related to the voltage are expected.

A time-dependent bias voltage $V(t)$ is a uniform potential in an electrical contact. This causes all the electronic wavefunctions to acquire a phase factor $e^{i\phi}$ with \cite{Tien}:
\begin{equation}
\phi(t)=\frac{e}{\hbar} \int_0^t V(t')dt'
\label{eq_TG}
\end{equation}
Eq.~(\ref{eq_TG}) captures the effect of both dc voltage and ac excitation on the transport. The dc voltage $V_{dc}$ leads to a linear increase of the phase with time, $\phi(t)=eV_{dc}t/\hbar$, which is equivalent to an energy shift by $eV_{dc}$, i.e. a shift of the chemical potential. The ac part of the voltage leads to a spreading in energy. For a linear system this is irrelevant, hence the ac voltage has no effect on dc average transport: there is no photo-assisted dc current. In contrast, the variance of current fluctuations (usually called noise) is a two-particle quantity for which the phase of the wavefunction matters. As a consequence, shot noise is sensitive to the ac voltage, a phenomenon called photo-assisted noise \cite{BuBlan,Nazarov_book}. This has been studied both theoretically and experimentally, for an ac excitation being a simple sine wave \cite{Lesovik,Rob,Reydellet,GR1}, as well as for more complex time-dependent periodic shapes \cite{Vanevic1,Vanevic2,Vanevic3,shaping} or non-periodic pulses \cite{Levitov,Ivanov}.

In this letter we report the observation of quantum oscillations of the low frequency excess noise of a a tunnel junction between normal metals excited by a time-dependent voltage $V(t)=V(1+\cos2\pi\nu t)$, as first proposed in \cite{Levitov}. The oscillations as a function of the voltage $V$ have a period $h\nu/e$. The experimental conditions to observe these oscillations are: i) a very low electron temperature $T=27$~mK, ii) the measurement of the noise generated by the sample (a tunnel junction) at low frequency, here in the range 1-80~MHz ($\ll k_BT/h=440$~MHz), and iii) an excitation at very high frequency, here 10 or 20~GHz ($\gg k_BT/h$).

\begin{figure}[setup]
\begin{center}
\includegraphics[width=0.9\linewidth]{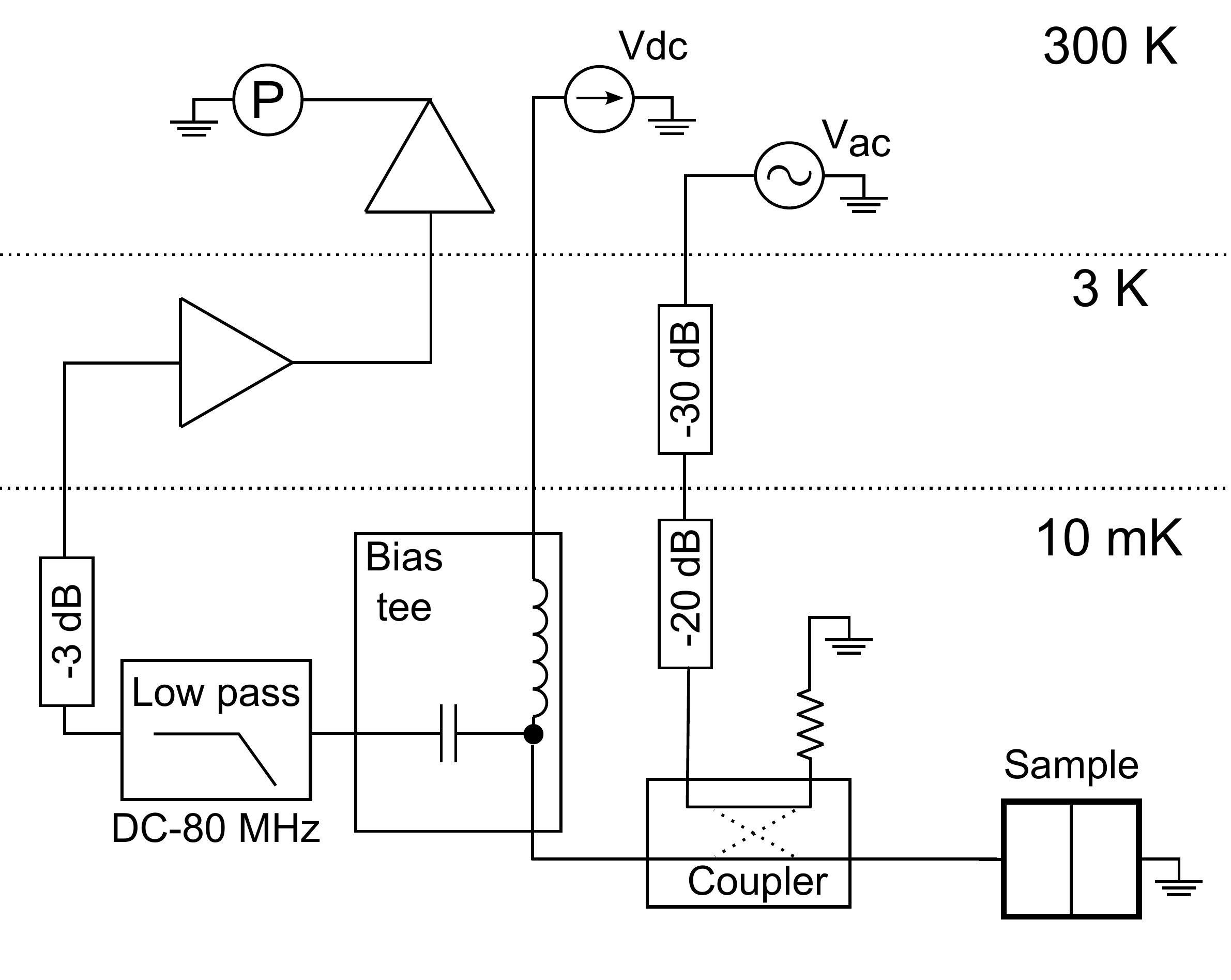}
\end{center}
\caption{Experimental setup. The \textcircledP \: symbol represents a power detector.}
\label{fig:setup}
\end{figure}

\begin{figure}
\centering
\includegraphics[width=1.1\columnwidth]{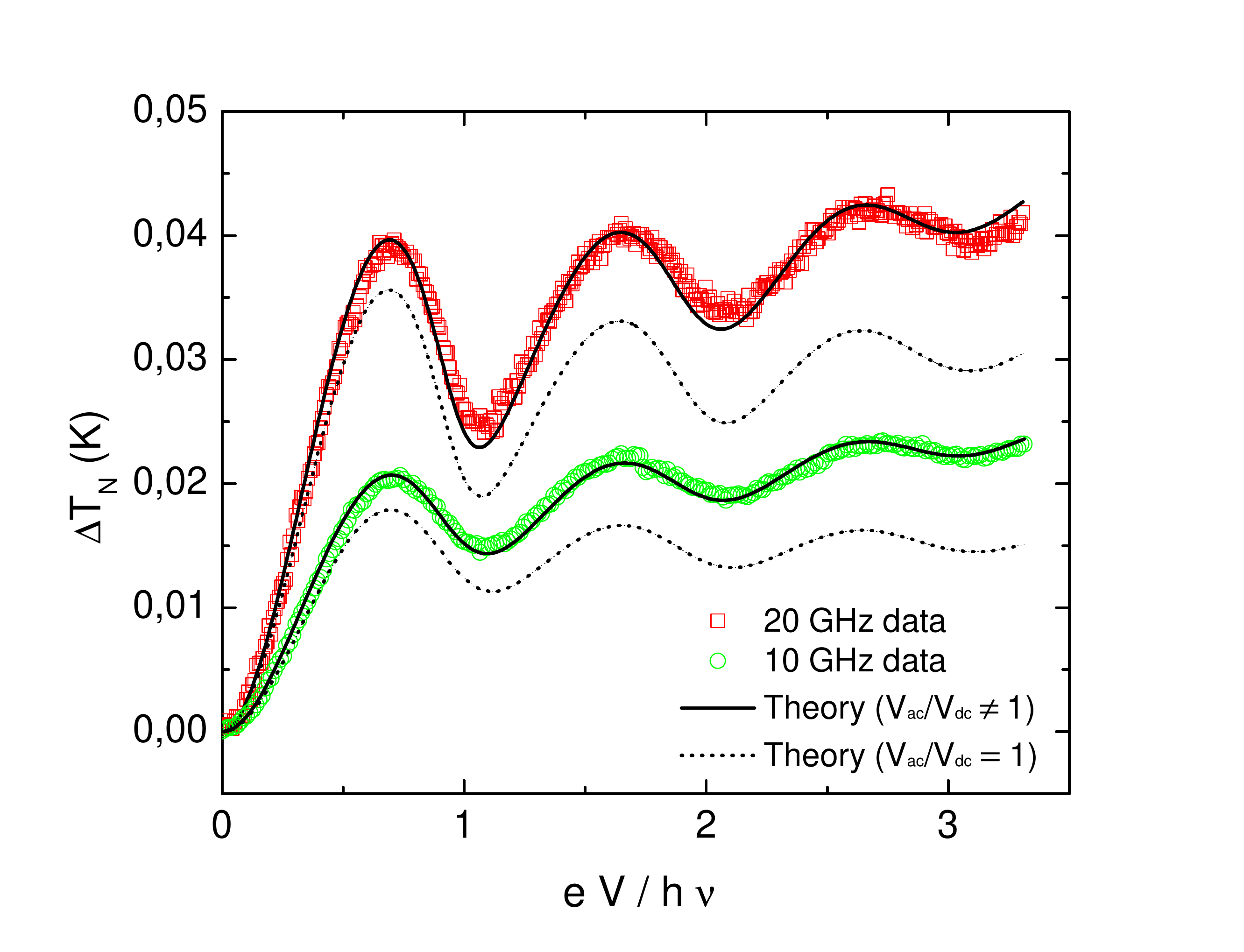}
\caption{(color online) Reduced excess noise $\Delta T_N = R\Delta S(V)/(2k_B)$ as a function of the amplitude $V$ of the time-dependent voltage excitation $V(1+\cos2\pi\nu t)$ applied on the sample. Symbols: experimental data for $\nu=20$~GHz (red square) and $\nu=10$~GHz (green circle). Dashed lines: theoretical expectations. Full lines: theoretical expectations for $V_{ac}/V_{dc}=1.075$ at 10GHz and 1.055 at 20 GHz.
}
\label{fig:excess_noise}
\end{figure}

\emph{Experimental setup.}
We cooled to 10~mK in a dilution refrigerator an Al/ Al oxide/ Al tunnel junction of resistance $R=70$~$\Omega$, similar to that used for noise thermometry \cite{Lafe}. A strong permanent magnet placed underneath the sample allowed to keep the aluminum normal even at the lowest temperatures. The detection setup is sketched on Fig.~\ref{fig:setup}. The tunnel junction is dc biased through a bias tee while the ac voltage fluctuations it generates are bandpass filtered in the range 1-80~MHz and amplified. A power detector measures the total power generated by the sample in that frequency range, which is given by $P=2GR_0S_IB$ where $G$ is the gain of the setup (amplifier, filter, cables), $B=79$~MHz the bandwidth, $R_0=50$~$\Omega$ the input impedance of the amplifier and $S_I$ the spectral density of current fluctuations at the input of the amplifier (the factor two stands for positive and negative frequencies). $S_I$ contains the contribution of the amplifier $S_a$ and that of the sample $S$, according to $S_I=S_a+(1-\Gamma^2)S$ with $\Gamma=(R-R_0)/(R+R_0)=0.29$ the reflection coefficient of the sample. All these quantities are almost frequency independent in the range 1-100~MHz. The noise spectral density $S(f)$ at frequency $f$ is related to the Fourier component $i(f)$ at frequency $f$ of the fluctuating current $i(t)$ by $S(f)=\langle |i(f)|^2\rangle$. The average $\langle . \rangle$ is performed experimentally by averaging over time. For example, each point in Fig.~\ref{fig:excess_noise} is averaged during 100~s. The microwave excitation of the sample takes place through a 16~dB directional coupler placed between the sample and the bias tee, since we could not find a bias tee that is good enough both below 100~MHz and at 20~GHz. A lot of attenuation ($50+16$~dB total) at various temperature has been inserted in the excitation line in order to avoid heating of the electrons by the room temperature radiation.

\emph{Calibration.}
We first consider the noise measured without ac excitation. We express the noise in terms of noise temperature referred to the input of the sample :

\begin{equation}
T_N=\frac{RS_I}{2k_B(1-\Gamma^2)}=\frac{RS}{2k_B}+\frac{T_a}{1-\Gamma^2}
\end{equation}

\begin{figure}
\centering
\includegraphics[width=1.1\columnwidth]{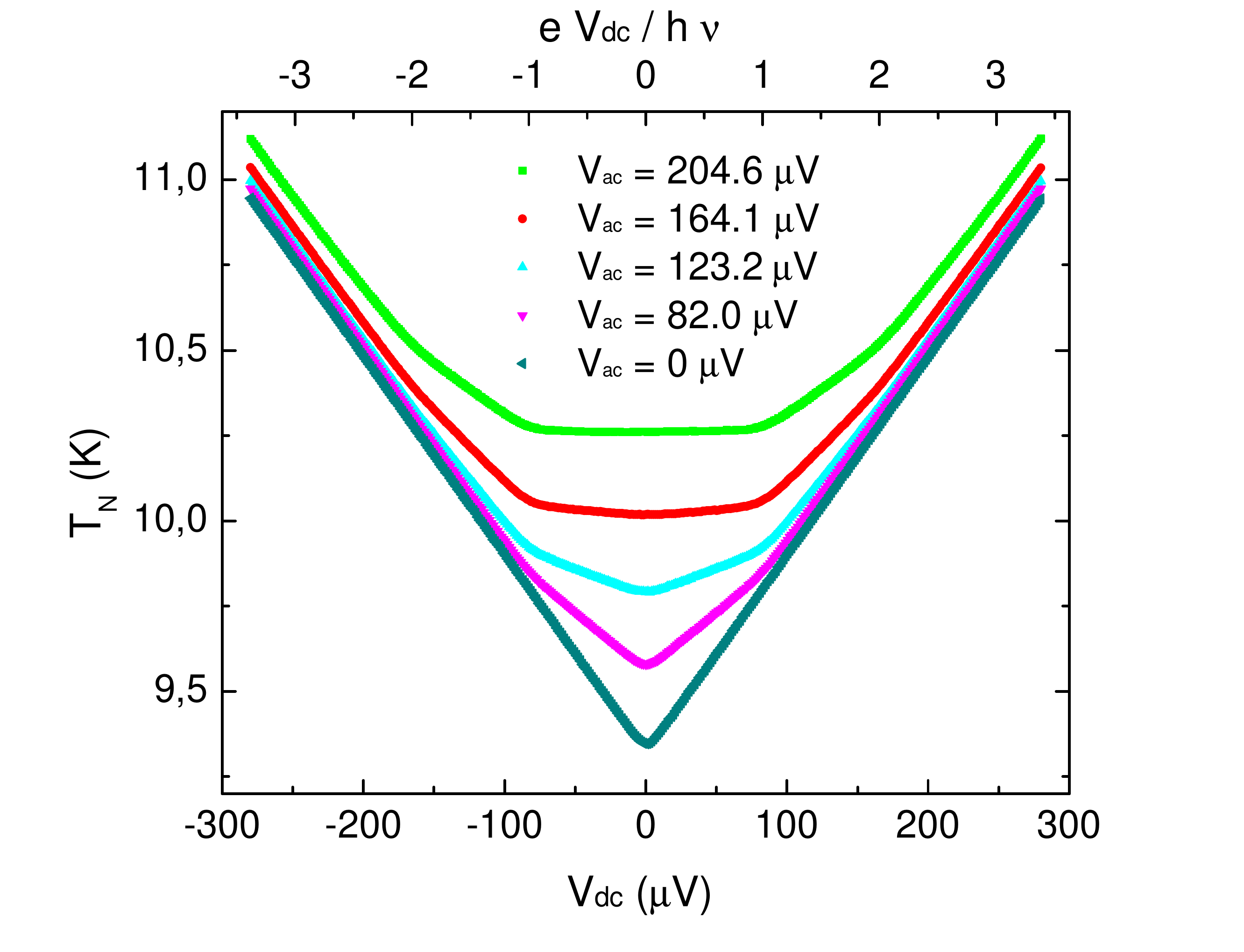}
\caption{(color online) Measured noise temperature as a function of dc voltage (lower axis) or reduced dc voltage $eV_{dc}/h\nu$ (upper axis) for various amplitudes of the ac excitation at frequency $\nu=20$~GHz. The theoretical fit, Eq.~(\ref{eq_PAN}), is indistinguishable from the experimental data.}
\label{fig:bruit_RF}
\end{figure}

The low frequency shot noise of the tunnel junction dc biased at voltage $V_{dc}$ is given by $S_0(V_{dc})=eR^{-1}V_{dc}\coth(eV_{dc}/2k_BT)$. By fitting the experimental result (the bottom curve on Fig.~\ref{fig:bruit_RF}) with this expression, we deduce the gain of the setup, the noise temperature of the amplifier $T_a=8.5$~K and the temperature of the electrons $T=27$~mK.

In the presence of an excitation at frequency $\nu$ and amplitude $V_{ac}$, the zero frequency shot noise becomes\cite{Lesovik}:

\begin{equation}
\label{eq_PAN}
S(V_{dc},V_{ac})= \sum^\infty_{n = -\infty} J^2_n\left( \frac{e V_{ac}}{h \nu} \right) S_0\left(V_{dc}+n \frac{h \nu}{e} \right)
\end{equation}
where $J_n$ are the Bessel functions of the first kind. We have measured photo-assisted noise vs. $V_{dc}$ for various ac excitations at frequencies 10~GHz and 20~GHz. We show on Fig.~\ref{fig:bruit_RF} our results for $\nu=20$~GHz. Shot noise in a quantum conductor under high frequency irradiation has already been measured both at low frequency \cite{Rob,Reydellet} and high frequency \cite{GR1}. However, no experiment had shown so clearly the structure of the experimental data: a piecewise linear increase of $S$ vs. $V_{dc}$ separated by kinks located at $V_{dc}$ multiple of $h\nu/e$, with the temperature appearing only as a rounding of each kink. This structure is a direct manifestation of the opening of inelastic channels for photo-assisted transport. In the absence of excitation and at zero temperature, only electrons of energy between $0$ and $eV$ can cross the junction. The ac excitation allows for absorption/stimulated emission of photons of energy $h\nu$. For $eV<h\nu$ no stimulated emission is possible. As soon as $eV>h\nu$, the electron can cross the sample without emitting a photon or it can emit one photon. At $2eV/h$  it can emit zero, one or two photons, etc. (and similarly for absorption) \cite{Lesovik}. This structure allows us to make a very precise calibration (within 1.5\%) of the ac voltage experienced by the sample: at dc voltage $V_{dc}=h\nu/2e$, the shot noise is hardly sensitive to the electrons temperature, as long as $k_BT\ll h\nu$. Thus we have measured $S(V_{dc}=h\nu/2,V_{ac})$ as a function of the excitation power and fitted the result to obtain $V_{ac}$. This avoids possible electron heating to affect the calibration of $V_{ac}$. Another issue to address is the heating of the electrons by the microwave power \cite{Reydellet}. In order to measure the electrons temperature in the presence of the microwave field, we have analyzed carefully the roundings of the measured noise as a function of $V_{dc}$ for fixed $V_{ac}$. It appears that for a given ac power, the kinks in $S$ can be well fitted by Eq.~(\ref{eq_PAN}) with a single temperature, i.e. $T$ is independent of $V_{dc}$. In contrast $T$ increases when $V_{ac}$ increases. We have extracted $T(V_{ac})$ by measuring the photo-assisted shot noise close to $V_{dc}=h\nu/e$, where the sensitivity to the temperature is maximal. We observe that $T$ increases from $27$~mK at $V_{ac}=0$ to $90$~mK for $V_{ac}=260~\mu$V at 20~GHz, which corresponds to $eV_{ac}/h\nu$ ranging from 0 to 3. We obtain comparable temperatures for the same ac voltage at 10 GHz, which corresponds to $T=65$~mK for $eV_{ac}/h\nu=3$ at that frequency.

\emph{Results.}
To reveal the quantum oscillations, we excite the sample with a time-dependent voltage  $V(t)=V(1+\cos2\pi\nu t)$, i.e. we sweep both the dc and ac voltages keeping  $V_{ac}=V_{dc}$. We measure the excess shot noise as a function of $V$, i.e.  $\Delta S(V)= S(V_{dc}=V_{ac}=V) - S(V_{dc}=V,V_{ac}=0)$. The amplitude of the oscillations $\sim10$~mK is very small compared to the drifts of the setup during the few hours of data acquisition. In order to minimize this effect, we have pulsed the RF excitation at 100~Hz while the modulation of the detected power, proportional to $\Delta S$, is measured with lock-in technique. The results for $\nu=20$~GHz and $\nu=10$~GHz as a function of $eV/h\nu$ are shown as symbols on Fig.~\ref{fig:excess_noise}. Oscillations of $\Delta S(V)$ with a period $h\nu/e$ are clearly visible on the experimental data. Note that the oscillations are not exactly periodic, as expected at finite temperature. When $T$ increases, the minima of $\Delta S$ are moved to higher voltage.

The theoretical expectations for $\Delta S(V)$ with a temperature that depends on $V_{ac}$ as obtained experimentally are presented on Fig.~\ref{fig:excess_noise} as dashed black lines. The full black lines on Fig.~\ref{fig:excess_noise} correspond to the theoretical predictions for $V_{ac}/V_{dc}=1.075$ instead of 1 at 10~GHz and $V_{ac}/V_{dc}=1.055$ at 20~GHz. These two curves reproduce our observations very well. However, despite many experimental checks (non-linearities in the detection, in the source, in the resistance of the sample) we do not have any explanation why our calibration would be off by 5-8\%. Note that at the frequencies we are working, the geometrical capacitance of the junction has a strong contribution to its impedance, and this may have consequences beyond a simple effective attenuation of the incoming excitation. In the same spirit, we have considered a $V_{ac}$-dependent electron temperature. This may be an oversimplification of the contacts of the junction being driven out of equilibrium by the ac excitation. And one may need to observe faint signals sensitive to details of the experiment, such as the present quantum oscillations, to address those issues.

\emph{Conclusions.}
We have shown that the low frequency shot noise of a tunnel junction driven out of equilibrium by a time-dependent voltage exhibits quantum oscillations. These oscillations are due to the interference in the transport of electron-hole pairs created at different times. They constitute a clear experimental demonstration of interferences in time domain, analogous to a multiple-slit interference in optics, generic of driven quantum systems \cite{Akkermans}. More experiments with more complex, time-dependent excitation or detection schemes could be envisioned in the same spirit, leading to a deeper understanding of the time-dependent transport in quantum systems.

We acknowledge fruitful discussions with W. Belzig and J. Gabelli. This work was supported by the Canada Excellence Research Chairs program, the MDEIE, the FCI, the NSERC, the FRQNT, the INTRIQ and the Canada Foundation for Innovation.


\end{document}